\documentclass[runningheads]{llncs}
\usepackage{graphicx}
\usepackage{multirow}
\usepackage[colorlinks,linkcolor=red, anchorcolor=blue, citecolor=blue, urlcolor=blue]{hyperref}
\usepackage{framed,multirow}
\usepackage{threeparttable}
\usepackage[misc]{ifsym}
\usepackage[symbol]{footmisc}
\begin{document}
\title{Rethinking the Extraction and Interaction of Multi-Scale Features for Vessel Segmentation}
\titlerunning{PC-Net for Vessel Segmentation}
\author{Yicheng Wu\inst{1,*} \and  Chengwei Pan \inst{2,*} \and Shuqi Wang \inst{3} \and Ming Zhang \inst{2} \and \\ Yong Xia \textsuperscript{1(\Letter)}\and Yizhou Yu \inst{3}}
\authorrunning{Yicheng Wu et al.}

\institute{
\textsuperscript{1} School of Computer Science and Engineering, \\Northwestern Polytechnical University, Xi’an 710072, China \\
\textsuperscript{2} Department of Computer Science, School of EECS, \\Peking University, Beijing 100871, China\\
\textsuperscript{3} Deepwise AI Lab, Beijing 100080, China\\
\textsuperscript{*} Equal Contribution \\
\email{yxia@nwpu.edu.cn}}
\maketitle 
\begin{abstract}
Analyzing the morphological attributes of blood vessels plays a critical role in the computer-aided diagnosis of many cardiovascular and ophthalmologic diseases. Although being extensively studied, segmentation of blood vessels, particularly thin vessels and capillaries, remains challenging mainly due to the lack of an effective interaction between local and global features. 
In this paper, we propose a novel deep learning model called PC-Net to segment retinal vessels and major arteries in 2D fundus image and 3D computed tomography angiography (CTA) scans, respectively.
In PC-Net, the pyramid squeeze-and-excitation (PSE) module introduces spatial information to each convolutional block, boosting its ability to extract more effective multi-scale features, and the coarse-to-fine (CF) module replaces the conventional decoder to enhance the details of thin vessels and process hard-to-classify pixels again.
We evaluated our PC-Net on the Digital Retinal Images for Vessel Extraction (DRIVE) database and an in-house 3D major artery (3MA) database against several recent methods. Our results not only demonstrate the effectiveness of the proposed PSE module and CF module, but also suggest that our proposed PC-Net sets new state of the art in the segmentation of retinal vessels (AUC: 98.31\%) and major arteries (AUC: 98.35\%) on both databases, respectively.

\end{abstract}

\section{Introduction}
The morphological attributes of blood vessels can be used for centerline extraction \cite{centerline}, image registration \cite{zheng} and artery/vein classification \cite{zhao}. Therefore, vessel segmentation is an essential step to construct a computer aided diagnosis system so that it is possible to do the large-scale screening of cardiovascular and ophthalmologic diseases \cite{fraz}.

Recently, many vessel segmentation methods have been published in the literature \cite{fu,yan}. Wu et al. \cite{wu1} proposed a multi-scale network followed network model to segment retinal vessels on fundus images. To address the issue of limited training data, they further introduced a deep supervision strategy to train the convolutional neural network \cite{wu2}. Zhang et al. \cite{yishuo} employed additional labels to reduce the intra-vessel differences. The channel-wise attention mechanism has also been employed to further improve the performance of retinal vessel segmentation \cite{dual}. Yu et al. \cite{yu} transferred the knowledge from retinal vessels to cardiac vessels for annotation-free segmentation. Furthermore, to improve the topology coherence, Araújo et al. \cite{topo} presented a cascaded model with variational auto-encoder (VAE) to maintain the spatial structures of vessels. Liu et al. \cite{liu} proposed an unsupervised ensemble strategy to combine the results produced by multiple methods.

Most of these methods are based on the U-shaped architecture \cite{unet}, which has achieved proven performance in medical image segmentation. However, the pooling layer inside U-shaped models would actually destroy the details of small objects. Even using the skip connections to transfer features from the encoder to decoder, the global information still dominates the segmentation process due to more channels containing global features. As a result, U-shaped models usually lack an effective interaction between local and global features.

In this paper, we propose a novel deep model called PC-Net to achieve better extraction and interaction of multi-scale features for vessel segmentation. We first incorporate spatial information into the squeeze-and-excitation (SE) module via dividing the image into pyramid blocks, and consequently construct the pyramid squeeze-and-excitation (PSE) module, which is then embedded into all convolutional blocks of our PC-Net. Moreover, we design a coarse-to-fine (CF) module to replace the conventional decoder of U-Net, aiming to relieve the side-effects of pooling layers. In the CF module, we first generate the residual feature maps, and then concatenate them with the feature maps produced by the conventional decoder. Thus, we can enhance the details of vessels and handle hard-to-classify pixels again. We evaluated our PC-Net on the Digital Retinal Images for Vessel Extraction (DRIVE) database and an in-house 3D major artery (3MA) database against several state-of-the-art methods and achieved superior performance over them.

The contributions of this work are three-fold: (1) designing the PSE module to extract more effective multi-scale features, (2) designing the CF module to refine the interaction of multi-scale features and enhance vessel details, and (3) setting new state of the art in the segmentation of retinal vessels and major arteries on the DRIVE and 3MA databases, respectively.

\section{Database}
We evaluated our proposed PC-Net on the DRIVE database \cite{staal} and the 3MA database.
The DRIVE database contains 40 2D color fundus images of retinal vessels with a size of $584\times565$, including 20 images for training and the rest 20 images for testing.
The 3MA database contains 100 computed tomography angiography (CTA) scans of major arteries in the head and neck captured by TOSHIBA CT machine. Each scan has a size of $512\times512\times D$ and a spatial resolution of $0.586\times0.586\times0.80mm^{3}$, where $D \in \left[369, 476 \right]$ is the number of slices. The first 50 scans are used for training, the next 20 scans for validation, and the rest 30 scans for testing. Each scan is equipped with a manual annotation of major arteries, which were first labeled by a radiologist and then verified by a senior radiologist.

\section{Method}
\begin{figure*}[htb]
\centering
\includegraphics[width=1\textwidth]{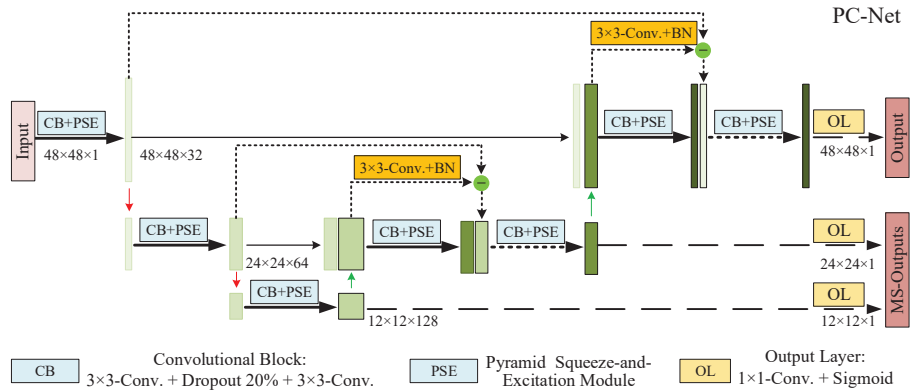}
\caption{\label{PCNet}Diagram of our proposed PC-Net}
\end{figure*}
The proposed PC-Net is designed based on U-Net\cite{unet} and has two unique features. We embed the PSE module into each of seven convolutional blocks and employ the CF module as the decoder to improve the performance of vessel segmentation. The diagram of our PC-Net is shown in Fig.~\ref{PCNet}.
\subsection{PSE Module}
\begin{figure*}[htb]
\centering
\includegraphics[width=1\textwidth]{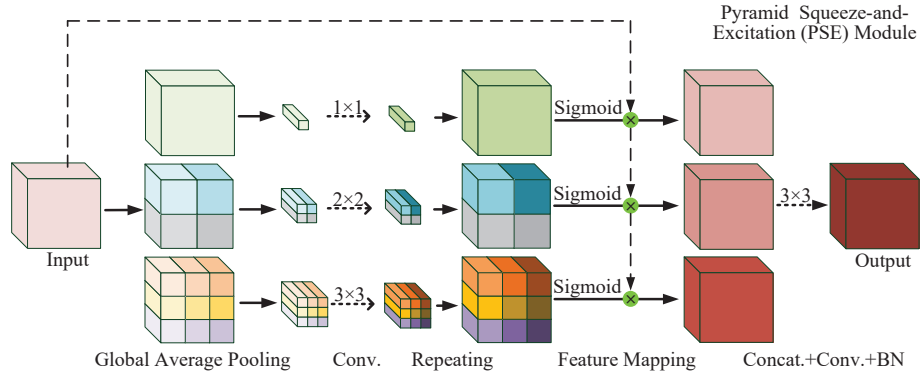}
\caption{\label{PSE}Diagram of PSE module}
\end{figure*}
Self-attention mechanism has a widespread application to computer vision tasks. The non-local neural network \cite{Non-local} focuses on the similarities of pixels. The SE-Net \cite{se} estimates and uses the importance of channels. Both DA-Net \cite{DA} and GC-Net \cite{gc} generate the attention maps based on channel-wise and spatial strategies. These models, however, have huge computational complexity due to the calculations among all pixels.

Thus, we introduce the spatial information into the SE module, and thus propose the PSE module, which is a more effective self-attention module (See Fig.~\ref{PSE}). We first divide each input feature map into regular blocks. Then we insert the global average pooling in each block and generate $1\times1$, $2\times2$ and $3\times3$ features. Next, we apply the convolutions of corresponding kernels to multi-scale features, and generate three sets of channel-wise weights after using the Sigmoid function. To map the weights, we further perform the channel-wise multiplication between weights and input features as \cite{se}. Finally, we combine all obtained features via a concatenate operation, a convolutional layer with $3\times3$ kernels and a batch normalization layer.

\subsection{CF Module}
\begin{figure*}[htb]
\centering
\includegraphics[width=1\textwidth]{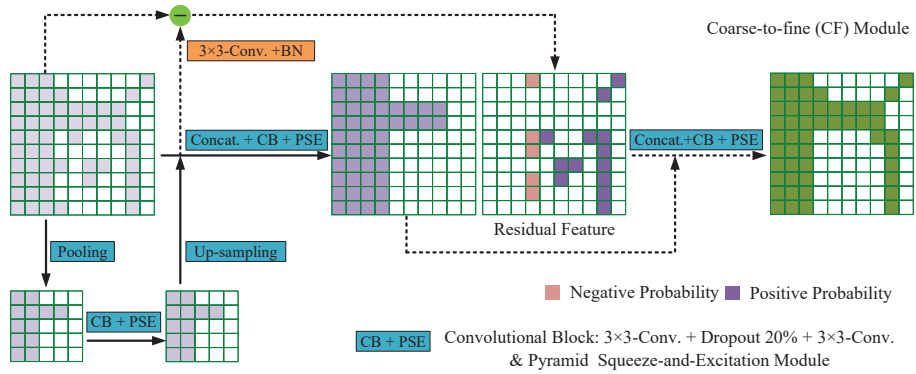}
\caption{\label{CF}Example of CF module}
\end{figure*}
The bilinear interpolation is usually used to expand the size of feature maps especially in the U-shaped model. Nevertheless, this simple up-sampling operation would eliminate some fine structures of tubular vessels. Therefore, the skip connections from encoder to decoder are inserted into U-Net \cite{unet} to relieve such negative impacts. However, this strategy can not change the domination of global information since the majority of channels in the concatenated feature maps are still up-sampled features.

To detect those hard-to-classify pixels located on the edge of vessels and capillaries, we design the CF module (See Fig.~\ref{CF}) and process these areas again. We first reduce the dimensions of expanded features using a $3\times3$ convolutional layer and a batch normalization layer, and then subtract the expanded features from the features produced by the front encoder. 

Thus, we obtain the residual feature maps, in which positive values mean the possibility of under-segmented pixels and negative values mean the possibility of over-segmented pixels. Next, we concatenate the residual feature maps with the feature maps produced by the conventional decoder, and then perform the convolutions again. With the help of residual maps, the decoder is, therefore, able to learn the class labels of hard pixels twice, leading to improved vessel details in the segmentation result.

\subsection{Loss Function} 
Besides the conventional output, our PC-Net also generates two multi-scale outputs (see Fig.~\ref{PCNet}). Therefore, the proposed PC-Net can be trained by minimizing the following weighted sum of the conventional binary cross-entropy loss and two multi-scale auxiliary losses
\begin{equation}
 Loss= CE(Y_{pred}, Y_{true})+\sum_{i=2}^{3} \lambda_i \times CE(msY_{predi}, msY_{truei})
\end{equation}
where $CE$ represents the binary cross-entropy function, $Y_{pred}$ is the predicted probability map of size $48\times48$, $Y_{true}$, $msY_{true2}$ and $msY_{true3}$ denote the ground truth of size $48\times48$, $24\times24$ and $12\times12$, respectively, and $\lambda_2$ and $\lambda_3$ are weighting parameters. The multi-scale ground truth $msY_{true2}$ and $msY_{true3}$ are obtained by down-sampling $Y_{true}$ via $2\times2$ and $4\times4$ max-pooling, respectively.

\subsection{Implementation Details}
We leveraged the max-pooling layer for down-sampling and the bilinear interpolation for up-sampling. In the output layer, we employed a $1\times1$ convolutional layer to reduce the number of channels and used the Sigmoid function to generate the probability output. Except for that, all activation functions were set to ReLU and the Adam optimizer was used for stochastic gradient descent with a learning rate of 0.001. To balance three multi-scale outputs, we empirically set the weighting parameter $\lambda_2$ to 0.67 and weighting parameter $\lambda_3$ to 0.33.

On the DRIVE database, we first employed the CLAHE algorithm \cite{clahe} and gamma adjusting method to pre-process each fundus image, and then adopted the data augmentation techniques including rotation and flip to increase the training data. Next, we randomly extracted 500k retinal patches of size $48\times48$ on the original and augmented fundus images for training. The batch size was set to 64 for 2D retinal patches.

On the 3MA database, we first clipped voxel values to the range of [0, 900] hounsfield units, and then linearly mapped voxel values to [0, 1]. Statistically, only 0.43\% voxels belong to vessels on each CTA scan. Due to the serious imbalance between vessel and background, we extracted 105 patches of size $48\times48\times48$ along vessels and 17 patches of the same size in the background of each scan. Thus, we extracted totally 6100 patches from 50 training scans. The batch size was set to 12 because of the limited GPU memory. Note that, all layers inside our PC-Net were replaced by corresponding layers of 3D kernel to enable our model to process 3D CTA patches. We further removed the isolated objects whose area is less than 40 voxels as post-processing to suppress the impact of small artifacts.

\section{Results}
\begin{figure*}[htb]
\centering
\includegraphics[width=1\textwidth]{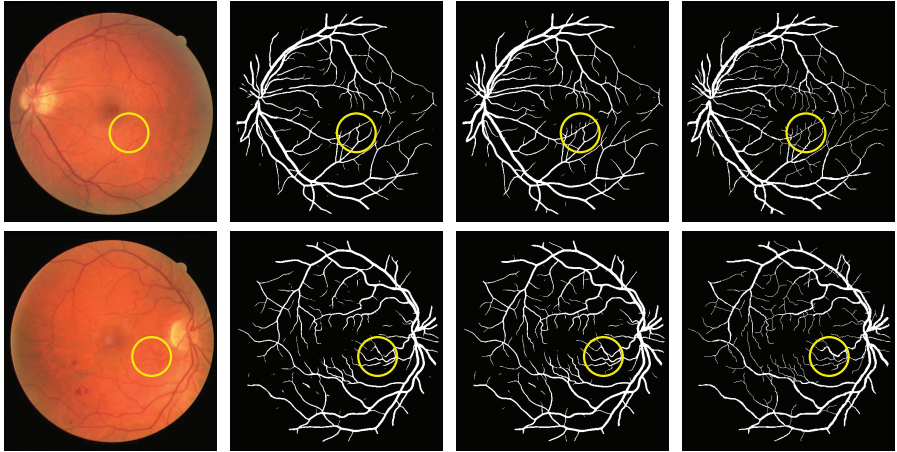}
\caption{\label{result1} Two fundus images in the DRIVE database (1st column), the segmented results obtained by using U-Net (2nd column) and our proposed PC-Net (3rd column), and the ground truth (4th column).}
\end{figure*}
\noindent\textbf{Results on DRIVE database:} Fig.~\ref{result1} shows two fundus images in the DRIVE database, the segmented results obtained by using U-Net and our PC-Net, and the ground truth. It reveals that our PC-Net is able to detect most of thin vessels and maintain the complex spatial structures (highlighted in yellow circles).

Table~\ref{tab:tab1} gives the retinal vessel segmentation performance of a human observer, six recently published methods and our PC-Net on the DRIVE database, which was measured in terms of the area under ROC curve (AUC), accuracy (Acc), specificity (Sp), sensitivity (Se). 
It shows that our PC-Net beats the human observer in all performance metrics with a substantial margin. Comparing to six existing methods, our PC-Net achieves the highest AUC of 98.31\%, highest Acc of 95.85\%, slightly lower Sp than \cite{wu1} and lower Se than \cite{topo}, \cite{yishuo} and \cite{liu}. Since there is a trade-off between Sp and Se, the highest AUC, which is not influenced by the binary threshold, and Acc indicate that our PC-Net outperforms these existing retinal vessel segmentation methods.

{
\begin{table*}[!htbp]
	\centering
	\caption{Performance of six recent segmentation methods and our PC-Net on the DRIVE database}
	\label{tab:tab1}
	\begin{center}
	\begin{tabular}{l|c|cccc}
		\hline 
		Methods&Year&AUC(\%)&Acc(\%)&Sp(\%)&Se(\%)\\
		\hline
		2nd observer &- &- &94.72 &97.24 &77.60\\
	    \hline
		Wu et al \cite{wu1}. &2018 &98.07&95.67&\textbf{98.19}&78.44\\   
		Zhang et al. \cite{yishuo}&2018  &97.99 &95.04 &96.18 &87.23\\
		Wu et al. \cite{wu2} &2019 &98.21 &95.78 &98.02 &80.38\\
		Liu et al. \cite{liu} &2019  &97.79&95.59&97.80&80.72\\
		Wang et al. \cite{dual} &2019 &97.72 &95.67 &98.16 &79.40\\
		Araújo et al. \cite{topo} &2019 &97.9 &- &95.3 &\textbf{89.7}\\
		\hline
		Our PC-Net &2020&\emph{\textbf{98.31}} &\emph{\textbf{95.85}} &\emph{98.09} &\emph{80.49}	\\
		\hline
	\end{tabular}
	\end{center}
\end{table*}
}
{
\begin{table*}[!htbp]
	\centering
	\caption{Ablation studies of our PC-Net on the DRIVE database}
	\label{tab:tab2}
	\begin{center}
	\begin{tabular}{l|ccccc|cc}
		\hline 
		Model(\%)&AUC&Acc&Sp&Se&Dice&Para.&FLOPs\\
	    \hline
		U-Net w/o DS&97.85
&95.58 &97.91 &79.57 &82.07 &0.47M &0.26G
\\
		U-Net&98.05 
&95.68 &98.08 &79.20 &82.34 &0.47M &0.26G
\\
		\hline
		U-Net with SE&98.21 
&95.79 &98.13 &79.76 &82.82 &0.48M &0.26G
\\
		U-Net with DA&98.17
&95.70 &98.40 &77.15 &82.04 &0.99M &2.06G
\\
		\hline
		U-Net with PSE&98.26
&95.81 &98.25 &79.05 &82.76 &1.24M &0.58G
\\
		U-Net with CF&98.26
&95.76 &\textbf{98.54} &76.68 &82.15 &0.70M &0.47G
\\
		Our PC-Net&\textbf{98.31}
&\textbf{95.85} &98.09 &\textbf{80.49} &\textbf{83.15} &1.62M &0.92G
\\
		\hline 
	\end{tabular}
	\end{center}
\end{table*}
}

To demonstrate the performance gain caused by each contribution, we carried out ablation studies using the same hyper-parameter setting. We chose the U-Net with deep supervision as the baseline and compared its performance to the performance of this baseline with various improvements. Besides the previously used performance metrics, we also reported the average Dice of segmentation results, and the number of parameters (Para.) and floating point operations per second (FLOPs) of each model in  Table~\ref{tab:tab2}, in which Para. and FLOPs measure the spatial and computational complexity of each model.
It shows that the deep supervision (DS) technique \cite{wu2} improves AUC by 0.20\%, the squeeze-and-excitation (SE) module increases the AUC by 0.16\%, and dual attention (DA) module results in a AUC gain of 0.12\%, either the PSE or CF module leads to a 0.21\% increase of AUC, and our PC-Net that uses both PSE and CF modules further improves AUC by 0.05\% over the model that uses either PSE or CF.
More important, our PC-Net achieves the best performance on the DRIVE database, better than the performance of U-Net with either the SE or DA module.

\begin{figure*}[!htb]
\centering
\includegraphics[width=1\textwidth]{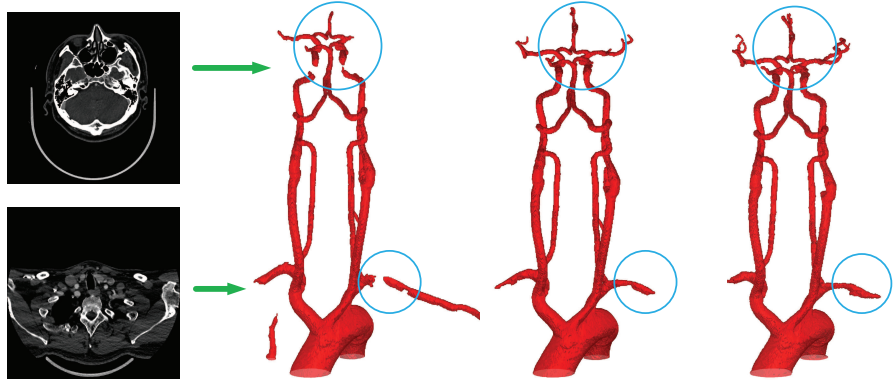}
\caption{\label{result2} Two CTA transverse scans in the 3MA database  (1st column), the segmented results obtained by using U-Net (2nd column) and our proposed PC-Net (3rd column), and ground truth (4th column)}
\end{figure*}

\noindent\textbf{Results on 3MA database:} Fig.~\ref{result2} shows two CTA transverse scans in the 3MA database, the results obtained by using U-Net and our PC-Net, and the ground truth. It reveals that the proposed PC-Net predicts more correct objects in 3D vessel segmentation. Specifically,
our PC-Net can extract most of intracranial vessels and does not over-segment many thick vessels (highlighted in blue circles).

We performed the same ablation studies again to show the performance gain caused by each contribution. Table~\ref{tab:tab3} gives the performance of all vessels and intracranial vessels on the 3MA database.
The results of all vessels indicate that the DS strategy \cite{wu2} improves AUC by 0.05\%, the SE module leads to a 0.07\% increase of AUC, and DA module results in a AUC gain of 0.29\%, and the designed PSE or CF module increases the AUC by 0.18\% or 0.33\%, respectively.
Then the proposed PC-Net, using both PSE and CF modules, further improves the performance and achieves the highest AUC of 98.35\%.
Moreover, the results of finer intracranial vessels also reveal that our PC-Net can accurately segment thin vessels.
As a consequence, the proposed PC-Net is able to achieve the best performance in 3D vessel segmentation. Note that, our PC-Net still outperforms U-Net model either with SE or DA module.

{
\begin{table*}[!htbp]
	\centering
	\caption{Performance of our proposed PC-Net on the 3MA database.}
	\label{tab:tab3}
    \begin{threeparttable}
	\resizebox{\textwidth}{!}{
	\begin{tabular}{l|ccccc|ccccc|cc}
		\hline 
		\multirow{2}{*}{Model}&\multicolumn{5}{c}{All Vessels(\%)}&\multicolumn{5}{|c}{Intracranial Vessels(\%)}&\multicolumn{2}{|c}{Complexity}\\
		\cline{2-13}
		&AUC&Acc&Sp&Se&Dice&AUC&Acc&Sp&Se&Dice&Para.&FLOPs\\
		\hline
		U-Net w/o DS
		&97.95 &99.91 &99.92 &95.97 &90.15
		&91.94 &99.98 &99.99 &83.90 &85.34 
		&1.41M &24.99G\\
		U-Net
		&98.00 &99.94 &99.95 &96.05 &93.40 
		&92.16 &99.98 &99.99 &84.32 &87.21 
		&1.41M &24.99G\\
		\hline
		U-Net with SE
		&98.07 &99.92 &99.93 &96.21 &91.26 
		&91.26 &99.97 &99.98 &82.53 &81.89 
		&1.42M &25.00G\\
		U-Net with DA\tnote{*}
		&98.29 &99.94 &99.95 &96.63 &93.19 
		&93.40 &99.98 &99.99 &86.81 &87.26 
		&2.54M &28.20G\\
		\hline
		U-Net with PSE
		&98.18 &99.94 &99.96 &96.39 &94.22
		&93.17 &99.98 &\textbf{100} &86.34 &89.90
		&3.69M &54.87G\\
		U-Net with CF
		&98.33 &99.94 &99.95 &96.71 &93.46 
 		&93.22 &99.98 &99.99 &86.45 &88.71 
		&2.10M &47.96G\\
		Our PC-Net
		&\textbf{98.35}&\textbf{99.95}&\textbf{99.96}&\textbf{96.75} &\textbf{94.46} & \textbf{93.63} &\textbf{99.98}&99.99&\textbf{87.27} &\textbf{89.97} 
		&4.82M &91.63G\\
		\hline 
	\end{tabular}}
    \begin{tablenotes}
    \footnotesize
    \item[*]DA module is applied to process $12\times12\times12$ features due to the limited memory
    \end{tablenotes}
    \end{threeparttable}
\end{table*}
}
\section{Conclusion}
In this paper, we present PC-Net for blood vessel segmentation in 2D and 3D medical images, which uses the PSE module to extract more effective multi-scale features and the CF module to enhance vessel details. Our comparative experimental results on the DRIVE database and 3MA database suggest the proposed PC-Net is able to deliver the state-of-the-art retinal vessel and major artery segmentation performance on both databases, respectively. Our ablation study also indicates the effectiveness of the PSE module and CF module. Future work will include encoding the spatial structures of vessel to improve the topology.

\end{document}